\documentclass[prl,aps]{revtex4}
\usepackage[english]{babel} 
\usepackage[protrusion=true,expansion=true]{microtype} 
\usepackage{amsmath,amsfonts,amsthm} 
\usepackage[svgnames]{xcolor} 

\usepackage{bm}
\usepackage[colorlinks=true,linkcolor=blue]{hyperref}
\usepackage{amsmath}
\usepackage{amssymb}
\usepackage{amsthm}
\usepackage{amsfonts}
\usepackage{times}
\usepackage{enumerate}
\usepackage{latexsym}
\usepackage{ifpdf}
\usepackage{graphicx}
\usepackage{color}
\usepackage{makeidx}
\usepackage{ulem}

\expandafter\ifx\csname package@font\endcsname\relax\else
\expandafter\expandafter
\expandafter\usepackage
\expandafter\expandafter
\expandafter{\csname package@font\endcsname}%
\fi
\begin{document}

\title{Competition between heavy-fermion and Kondo interaction in isoelectronic  A-site ordered perovskites}

\author{D. Meyers$^{1,*}$}
\author{S. Middey$^1$}
\author{J.-G. Cheng$^{2,3}$}
\author{Swarnakamal Mukherjee$^4$}
\author{B. A. Gray$^1$}
\author{Yanwei Cao$^1$}
\author{J.-S. Zhou$^3$}
\author{J. B. Goodenough$^3$}
\author{Yongseong Choi$^5$}
\author{D. Haskel$^5$}
\author{J. W. Freeland$^5$}
\author{T. Saha-Dasgupta$^{4,\dagger}$}
\author{J. Chakhalian$^{1}$}

\address{$^{1}$Department of Physics, University of Arkansas, Fayetteville, AR 72701, USA}
\address{$^{2}$Beijing National Laboratory for Condensed Matter Physics, and Institute of Physics, Chinese Academy of Sciences, Beijing 100190, China}
\address{$^{3}$Texas Materials Institute, ETC 9.102, University of Texas at Austin, Austin, Texas 78712, USA}
\address{$^{4}$Department of Condensed Matter Physics and Materials Science, S.N.Bose National Centre for Basic Sciences, Kolkata 700098, India}
\address{$^{5}$Advanced Photon Source, Argonne National Laboratory, Argonne, Illinois 60439, USA}
\address{$^{*}$ Correspondence to dmeyers@uark.edu, $^{\dagger}$ t.sahadasgupta@gmail.com}

\begin{abstract}
With current research efforts shifting towards the 4$d$ and 5$d$ transition metal oxides, understanding the evolution of the electronic and magnetic structure as one moves away from 3$d$ materials is of critical importance. Here we perform X-ray spectroscopy and electronic structure calculations on $A$-site ordered perovskites with Cu in the $A$-site and the $B$-sites descending along the 9th group of the periodic table to elucidate  the emerging properties as  $d$-orbitals change from partially filled 3$d$, 4$d$, to 5$d$. The results show that when descending from Co to Ir the charge transfers from the cuprate like Zhang-Rice state on Cu to the t$_{2g}$ orbital of  the B site. As the Cu $d$-orbital occupation approaches the Cu$^{2+}$ limit, a mixed-valence state in CaCu$_3$Rh$_4$O$_{12}$ and heavy fermion state in CaCu$_3$Ir$_4$O$_{12}$ are obtained. The investigated d-electron compounds are mapped onto the Doniach  phase diagram of the competing RKKY and Kondo interactions developed for \textit{f}-electron systems.
\end{abstract}

\maketitle

\clearpage
\newpage
Transition metal (TM) oxides host a diversity of fascinating phenomena\cite{Imada, Jak1, Jak2, Tokura, McCormack, Wang}. Several possible formal oxidation states of the TM ions coupled with  the innate ability to stabilize those states by structural network of oxygens give rise to a striking change in the TM -- O orbital hybridization $W$, electron-electron correlations $U$, and the charge transfer energy $\Delta$ \cite{Zaanen}; the subtle competition between $W, U$ and $\Delta$ across the 3d group of the periodic table is  then responsible for a vast landscape of interesting magnetic and electronic ground states\cite{Imada,Tokura, Tokura2}.  In traversing the periodic table within groups of the 3$d$ $\rightarrow $4$d$$ \rightarrow $5$d$ TM blocks, the additional degree of freedom, the spin-orbit (SO) interaction, $\lambda$ gets activated and is predicted to  foster a multitude of  exotic electronic and topological phases of correlated matter \cite{Haskel,Kim,Laguna,balents, Cao, Cao2,ash}.

While the vast majority of these compounds contain TM ions at the $B$-site of the $AB$O$_3$ perovskite structure, it is now possible to synthesize a new family of compounds with formula unit $(A$$A^{\prime}_3$$)B_4$O$_{12}$, Fig. 1(a), where the perovskite $A$-site is partially occupied by a TM ion, labeled $A^\prime$. Among this class of materials, $(A$Cu$_3$$)B_4$O$_{12}$ with Cu in the $A^{\prime}$ site have garnered considerable attention due to the presence of CuO$_4$ planes. Structurally, this family of compounds consists of two different magnetically active  sub-lattices of TM ions:   $B$O$_6$ octahedral units (forming a 3D  octahedral network as in the typical $AB$O$_3$ perovskite lattice) and the planar CuO$_4$ units  coupled with $B$O$_6$ units via an apical oxygen (see Fig. 1(a)). Depending on the choice of $B$-site ion, the materials exhibit exciting properties including  giant dielectric constant ($B=$ Ti), exotic ferromagnetism ($B$ = Ge, Sn, Fe), valence fluctuation ($B$ = V), Mott physics ($B$ = Ru), and inter-site charge order ($B$ = Fe) to name a few\cite{Downes, ccgeo_prb, Yamada2, vanadium, Hollmann, Chen, Yamada}. They are also of particular interest due to the preservation of the cubic lattice symmetry (\textit{IM$\overline{3}$}) despite large variations of the $B$-site ion including utilizing different TM series (3$d$ or 4$d$ or 5$d$) of the periodic table\cite{Xin}. This provides a unique platform to  investigate the emergence of the electronic and magnetic states; for example, in recent work on (CaCu$_3)$$B$$_4$O$_{12}$ ($B$ = Cr, Co)  it has been demonstrated that the Zhang-Rice quantum state  (essential for hole doped high Tc superconductivity) can be realized  in these compounds, despite the lack of the superconducting  ground state~\cite{Meyers, Mizokawa, Zhang}. 

Very recently, Cheng \textit{et al}. showed that this class of materials displays a crossover between the magnetic insulating and paramagnetic metallic states, depending on the Cu-O and $B$-O bond lengths\cite{ChengUn}. It was further suggested that  the CaCu$_3$$Ir_4$O$_{12}$ (CCIrO) compound with the bond-length within the crossover region possesses anomalous electronic and magnetic properties arising presumably due to the interaction between the localized Cu and itinerant Ir states\cite{Xin}. However, the mechanism by which the electronic structure transforms to create this emergent behavior in CCIrO is not  microscopically understood. In order to shed light on this interesting aspect, three \textit{isostructural} and \textit{isoelectronic} compounds whose $B$-site spans the 9th group of the periodic table, (CaCu$_3$$)B_4$O$_{12}$ ($B$ = Co, Rh, Ir) were synthesized. Within the proposed phase diagram of Cheng \textit{et al}. $B$ = Co occupies the paramagnetic metallic state while CCIrO is at the crossover region as reflected for instance in anomalous  d.c. transport  properties shown  in Fig. 1(b)\cite{Yamada, ChengUn}. 

In this paper, we investigate the electronic and magnetic structure of this new class of materials by a combination of resonant soft and hard x-ray absorption spectroscopy (XAS) on the Cu L-edge, O K-edge, and $B$-site L- and K-edges. Complementary first-principles calculations corroborate the experimental findings, providing microscopic understanding. Our results reveal that the unusual  physical  properties  of these compounds are microscopically controlled by  the degree of Cu  3$d$$_{\textrm{x$^{2}$-y$^{2}$}}$ orbital occupancy and the strength of the $B$-O covalency.

\section{Results}

\subsection{Spectroscopic results}

First,  we discuss  the evolution of the electronic structure of Cu. The Cu L$_3$ edge XAS are  presented in Fig. 2(a) (TFY available in supplementary Fig. 1). In accord with the  previously reported results~\cite{Meyers},  for CaCu$_3$Co$_4$O$_{12}$ (CCCoO) the main absorption line at 931.4 eV arises due to absorption from the $d^9$\underline{L} $\to$ \underline{c}$d^{10}$\underline{L} transition, where  \underline{L} stands for a ligand hole whereas  \underline{c} indicates a hole in the Cu 2$p$ core states. The low-energy shoulder around  930 eV arises from the $d^9$ $\to$  \underline{c}$d^{10}$ transition; the $d^9$ transition accounting for 10\% of the total peak area, indicating that there is a large Cu - O hybridization present in CCCoO. The line shape of  the Cu L$_3$ absorption edge for CaCu$_3$Rh$_4$O$_{12}$ (CCRhO), containing the 4$d$ element Rh at the $B$-site, also contains a lesser and yet still significant $d^9$\underline{L} contribution, giving a Cu valence of $\sim$ 2.6 +. The transport behavior for this compound is quite similar to CCCO and points to CCRhO also being paramagnetic metallic, Fig. 1(b). On the other hand, for CCIrO, the $d^9$\underline{L} state is no longer present and the L$_3$ line shape is  almost  analogous  to that recorded from purely ionic  $d^{9}$ Cu$^{2+}$ charge state, agreeing with an earlier report utilizing electron energy loss spectroscopy (EELS) \cite{Xin,Jak1,Jak2,Hu, Sarangi,Achkar}. This implies that  the hole is no longer interacting with the Cu \underline{c}$d^{10}$ final state, showing a significant reduction of the hole contribution to the hybridized orbital between Cu and O. Furthermore, the small multiplet split peak seen around $\sim$ 940 eV for all Cu L$_3$-edge spectra is due to a transition from the metastable 3$d^8$ (Cu$^{3+}$) to \underline{c}$d^9$ state; the small decrease in this 3$d^8$ peak spectral  weight in going from Co to Ir also indicates movement towards the ionic 3$d^9$ state of Cu \cite{Kaindl,Sarangi, Hu}. Taken as a whole, the Cu L$_3$-edge data directly  illustrate that in the presence of  4$d$ or 5$d$ orbitals it is energetically more favorable to transfer the hole  on to the $B$-site instead of stabilizing the Cu$^{3+}$ state, suggesting a strong change in covalency, as tabulated in Table 1.  

In the past, the electronic structure of Cu  has been extensively  studied in the context of  high $T_c$ superconducting cuprates, where it was also found that the reduction of  ligand hole weight on Cu causes a decrease of the pre-peak intensity around 528 eV in the O K-edge XAS spectrum~\cite{Chen2, Nucker, Nucker2, Kuiper}.  To track the movement of the hole in the present series of samples, we obtained  O K-edge XAS  spectra shown in Fig. 2(b), with the post-edge normalized to 1 (at 540 eV). As immediately seen, the decrease of the relative intensity of the pre-peak indeed scales with the  reduction of the ligand hole on Cu, implying a marked change in both bandwidth,  $W$, and the charge transfer gap,  $\Delta$.
The decrease of the pre-peak intensity can be rationalized  in terms of a decreasing availability of empty states as the hole concentration on O decreases in moving from Co to Rh to Ir. While the high energy shoulder on the Cu L$_3$ peak disappears entirely for CCIrO, the oxygen prepeak does not, which indicates the hybridization between the $B$-site and O is also significant, as Cu 2+ will not contribute to the prepeak. The results of the O K edge spectroscopy are thus in excellent agreement  with the Cu  L-edge observations. Another interesting  observation is that the energy separation between the pre-peak and the peak around 530 eV increases from  Co to Ir. The shift towards higher energy will  be discussed in detail  in the theory section  and is  attributed to the gradually increasing separation of the $B$-site bands (CCCoO) and both the $B$-site and Cu bands (CCRhO and CCIrO) from the O 2$p$ band. 

The movement of the hole away from Cu naturally implies a change in valency of the $B$-site ion. To  verify this proposition and  to corroborate the previous findings we performed XAS on the L and K edges of Ir and Rh respectively (Co was measured previously and found to be in the $\sim$ 3.25+ state, after subtraction of an impurity peak\cite{Meyers}). Moving to  the 4$d$ compound, Rh K-edge XAS spectra  have been simultaneously  collected from the CCRhO sample and a standard SrRhO$_3$ reference sample. Though the line shape from CCRhO shares similarities with SrRhO$_3$ (Rh$^{4+}$), the difference in energy at 80 \% of the normalized absorption was found to be $\sim$ 0.68 eV lower relative to SrRhO$_3$, Fig. 3(a)\cite{Zeng}. Based on this shift, and the shift of $\sim$ 1.9 eV between Rh 3+ and 4+ (See supplementary Fig. 2), we conclude that the Rh valence state is indeed strongly mixed between 3+ and 4+, with a value of 3.64+ ($\pm$ 0.1). In conjunction with the Cu and O soft XAS, the entire Rh data set strongly supports the notion that the hole still largely resides on the O anion, but spreads  to the hybridized orbital with Rh.
Finally, the Ir L$_3$ edge (2$p_{3/2}$ $\to$ 5$d$ transition) recorded from CCIrO and SrIr(4+)O$_3$ is shown in Fig. 3(b). As seen, the remarkably similar  line shape and the energy  peak position both confirm that Ir is in the 4+ state. The L$_3$ to L$_2$ branching ratio  (BR) was found to be 3.79 (analysis shown in supplementary Fig. 2)\cite{Haskel, Laguna}. This value is similar to that found in several other iridate compounds and signifies a large spin-orbit coupling (SOC) as expected for a 5d compound\cite{Haskel,Laguna,Kol,Clancy}. Overall, the Ir L edge data is in excellent  agreement with the  Cu L edge data stating the $d^9$ Cu$^{2+}$ ground state, with a much smaller $d^8$ contribution  compared to the Rh and Co compounds; thus, in the CCIrO compound the hole is now almost  entirely transferred from the Cu 3$d$ - O 2$p$ state to the Ir 5$d$ - O 2$p$ hybridized orbital. The obtained valences for the $B$-sites are also listed in Table 1.

\subsection{DFT results}

Calculating the spin-polarized electronic structure of the three compounds, as a common feature, we find that the Cu $d$, $B$ $d$ and O $p$ states are admixed, though the degree of admixture varies between the three compounds. Site projected partial  density of states (PDOS) is shown in supplementary Fig. 3 \cite{swarna}. A direct inspection of the plot for CCCoO reveals that the Cu $d_{x^{2}-y^{2}}$ - O ($p$) states, which are strongly admixed with Co e$_g$ (e$^{\sigma}_{g}$) states, are empty in both spin up- and spin down channels, thus lending strong support for the  Cu$^{3+}$ ($d^{9}\underline{L}$) valence in CCCoO. This yields a mixed valence of 3.25+  on Co and  the intermediate spin state with a magnetic moment of $\sim1.68$ $\mu_B$. Moving towards the CCRhO compound, the admixture between the Cu $d_{x^{2}-y^{2}}$ + O  state and Rh states becomes  markedly reduced compared to that of CCCoO. In this  compound, the calculated Cu valence is found to be mixed between 2+ and 3+ ($\sim$ 2.5+), with  Rh valence close to 3.6+. Unlike Co, magnetically Rh is found to be in  the low spin state with a  spin moment of 0.21 $\mu_B$ and largely  quenched orbital moment of 0.05 $\mu_B$. 
Finally, we consider the CCIrO. In sharp  contrast to  both  Co  and Rh, the  mixing of the Cu $d_{x^{2}-y^{2}}$ - O $p$ states and the Ir $d$ states is drastically reduced. This results in an almost pure Cu $d_{x^{2}-y^{2}}$ - O $p$ state  occupied in one spin channel and empty in another, implying a Cu$^{2+}$ valence  and nominal  Ir\(^{4+}\) valence. The spin moments at Ir and Cu sites are found to be 0.45 $\mu_B$ and 0.65 $\mu_B$ respectively, with a rather large spin moment of 0.12 $\mu_B$ on O, arising from strong hybridization with Ir. We note that ionic Ir$^{4+}$ is in the 5$d^{5}$ configuration and  has extensively been discussed  in view of the interplay of strong SO coupling and correlation physics\cite{Haskel,Kim,Laguna,balents, Cao, Cao2,ash}. The large BR of close to 3.8 signifies the presence of a large spin-orbit coupling at the Ir site, which is found to be common among many of the compounds of Ir 4+ in an octahedral environment, even when the compound is metallic as in IrO$_2$\cite{Clancy}. The orbital moment at the Ir site, calculated within the GGA+U+SO, turned out to be 0.12 $\mu_B$, smaller than the spin moment with $m_{orbital}/m_{spin}$ being 0.27. This is curious when compared to the values obtained in cases like BaIrO$_3$\cite{Laguna}. The structure and coupling mechanisms are, however, rather different between compounds like Sr$_2$IrO$_4$ or BaIrO$_3$ and the present one. In the former examples, magnetic interactions are one or two-dimensional Ir - Ir, while here they are Cu - Ir with three dimensional connectivity. The dominance of hybridization produces an additional induced spin moment at the Ir site due to the presence of the magnetic ion Cu, a behavior qualitatively similar to the case of La$_2$CoIrO$_6$, discussed in  recent literature\cite{Kol}. The presence of a significant SO interaction is found to mix up and down spins and destroys half-metallicity in CCIrO. We note here that inclusion of spin-orbit coupling allows for the non-collinear arrangement of spins. We have therefore carried out
spin-polarized calculations assuming collinear as well as non-collinear spin arrangements. The calculated 
magnetic moments at various atomic sites, obtained in non-collinear calculations, turn out to be very 
similar to that obtained assuming simple collinear arrangement of spins. The spin magnetic moments are found to be similar to that obtained from collinear 
calculations within 1-2\%, while the orbital magnetic moments are found to differ from that in collinear calculation by a maximum of 0.5\%, 
providing confidence in the general conclusion drawn from the electronic structure calculations on the various valence and spin-states, irrespective
of the assumed spin arrangements.

The evolution of the nominal valence of Cu from predominant 3+ ($d^{9}\underline{L}$) to 2+, as one moves from 3$d$ (Co) to 4$d$ (Rh) to 5$d$ (Ir) elements at the $B$-site, is controlled by mixing between $B$-site $d$ states and Cu $d_{x^{2}-y^{2}}$ - O $p$ states and can be vividly  visualized in the effective Wannier function plots shown in Fig. 4(a) (upper panel). As clearly  seen, the Wannier functions centered at the Cu site have the orbital character of $d_{x^{2}-y^{2}}$ symmetry, and the tail is shaped according to the symmetry of the orbitals mixed with  it. Specifically, moving from CCCoO to CCRhO to CCIrO, we find the weight at the tails centered at  the $B$-site (marked by a circle) progressively diminishes.

Microscopically, the nature of this peculiar unmixing / dehybridization effect between Cu-O and $B$-site in moving from 3$d$ to 4$d$ to 5$d$ element at the $B$-site can be further elucidated by considering the energy level positions of $B$ $d$, Cu $d$, and O $p$ states (see Fig. 4(a) ( bottom panel)). As  mentioned  above, the octahedral crystal field coupled with the trigonal distortion separates the $B$ $d$ states into doubly degenerate e$^{\sigma}_g$, e$^{\pi}_g$ and singly degenerate a$_{1g}$ ones,  while the square planar geometry of the CuO$_2$ plane breaks the Cu d states into Cu $d_{x^{2}-y^{2}}$ and the rest, with  Cu $d_{x^{2}-y^{2}}$ being of the highest energy. In progressing from CCCoO to CCRhO to CCIrO, the relative position of Cu $d_{x^{2}-y^{2}}$ with respect to O $p$ states increases, driven by the pushing down of O $p$ states due to the increased crystal field splitting (e$^{\sigma}_g$ - e$^{\pi}_g$/a$_{1g}$ splitting) at the $B$-site. This, in turn, makes the hybridization between the Cu sublattice and $B$ sublattice weaker and weaker since those ions communicate via the intervening oxygen. This highlights a key difference between CCIrO and CaCu$_3$Ru$_4$O$_{12}$, where for the later it was found that the suspected Kondo-like physics was unlikely due to a strong mixing of Cu with O\cite{Mizumaki, Tanaka}.  Similar to  that of high $T_c$ cuprates,  for CCCoO the O $p$ states are  positioned \textit{above} Cu $d$$_{x^{2}-y^{2}}$, placing Cu in to a \textit{negative charge transfer} regime which promotes a high-$T_c$ cuprate like d$^9$$^{}\underline{L}$ state akin to  the Zhang-Rice singlet state\cite{Meyers, Mizokawa, Zhang,neg-ch}. The progressive weakening of covalency between the $B$ sublattice and Cu-O sublattice as one moves from CCCoO to CCRhO to CCIrO, makes the spread of the effective Cu $d_{x^{2}-y^{2}}$ Wannier function (top panel of Figure 4(a)) in the case of Ir dramatically reduced compared to either Co or Rh.

\subsection{Unified Picture}

Finally, the element  resolved spectroscopic results combined with the \textit{ab-initio} calculations prompts us to  build a unified framework to  explain their emergent  physical  behavior. While an earlier study utilizing EELS  found small changes between 3-4-5 $d$ A-site ordered perovskites from different columns\cite{Xin},  our study reveals that upon ascending a column of the periodic table from Ir to Co,  the Cu 3$d_{x^2-y^{2}}$ orbital    occupation changes from   practically ionic 3$d_{x^2-y^2}^9$ ($S$=1/2) for CCIrO to the non-magnetic cuprate Zhang-Rice like state with  3$d^9$\underline{L} ($S=0$) for CCCoO, as supported by experimentally and theoretically deduced valences listed in Table 1. Along  with  it, these  localized and magnetically active Cu $d$-states in CCIrO shift   towards the Fermi surface demonstrating  a rapid change in hybridization compared to both CCRhO and  CCCoO.  On the opposite end, such  a drastic change in the  Cu orbital  occupation results in the \textit{mixed valence}  intermediate spin state of Co$^{3.25+}$,  mixed valence Rh $\sim$ 3.7+  but ionic Ir$^{4+}$ (5$d^5$). These findings  allow us to  place the three compounds under discussion  in the context of the Doniach  phase diagram depicted  in Fig. 4(b), where the fundamental  control parameter is the average occupation $\langle n \rangle$$_\textrm{Cu}$ of the $d_{x^2-y^2}$ orbital   modulated by the hybridization from  the strongly  mixed $B$-site  $d$- and O $p$-bands\cite{Yi, Sullow, Si}. In the modern version of the Doniach phase diagram interesting  physics involving heavy fermions manifests itself as a competition between the Kondo liquid and spin liquid behavior mediated by chemical doping, while  very little attention has been given to the mixed valency regime, particularly in $d$-electron systems and in the absence of doping\cite{Yoshida, Pandey, Feng, Yi, Sullow, Si}. In this framework the overall   ground state is   defined by the competition between RKKY\   type magnetic  exchange between magnetic  holes on Cu with the Kondo  screening by  conduction carries from the B-O sublattices. For CCIrO,  with a  $S$ = 1/2 $d$-hole   localized on Cu, the large magnetic exchange   is  comparable in strength with the Kondo screening, resulting  in the strongly enhanced effective mass observed with transport  and thermal measurements\cite{ChengUn}. Thus, Cheng \textit{et al.} firmly placed CCIrO into the heavy fermion regime I in Fig. 4(b) with the antiferromagnetic local moment short-range magnetism\cite{ChengUn,Xin}. In moving from Ir to Rh and Co,  the  Kondo  energy  scale begins to  gain due to  the collective hybridization of Cu d-holes into the   ZR singlets. With  the strong  reduction  in the  Cu  $d_{x^2-y^2}$ orbital occupation    both CCRhO\  and CCCoO enter the regime II\ of mixed valency  (or Kondo  liquid phase)\ in Fig. 4(b). Unlike  regime I, in the mixed-valence regime  quantum fluctuations between different electronic configurations are highly relevant;  in this regime,   the local electronic and magnetic structure of Kondo  centers (Cu) is defined by  the  redistribution of  electrons between  Cu d-states and electrons from the strongly hybridized d and p-states of Rh (Co) and O, i.e.  $\vert$3$d^{9}$, $S$ = 1/2$\rangle$ vs.   $ \vert 3d^{9}\underline{L}, S = 0\rangle$. The conjectured microscopic  framework that  links  the electronic and magnetic ground state of the $A$-site perovskites with  macroscopic behavior   opens a path in designing emergent   ordered phases with  heavy  fermion behaviour,  quantum criticality, and  unconventional superconductivity  in the  magnetic Kondo lattice of cuprate-like moments.  
 
To summarize, we performed XAS measurements and first-principles calculations on a series of $A$-site ordered perovskites, chemical formula CaCu$_3$$B$$_4$O$_{12}$, spanning one period of the periodic table. Surprisingly, we find that the materials fit well within the Doniach phase diagram, being controlled by the hole count on Cu, leading to the conclusion that the competition between RKKY and Kondo effects is responsible for the anomalous behavior observed in the CCIrO compound.

\section{Methods}

All samples used in the present study were prepared under high-pressure and high-temperature (HPHT) conditions with a Walker-type multianvil module (Rockland Research Co.). The A-site-ordered perovskites CCCoO, CCRhO, and CCIrO were obtained under P = 9 GPa and T = 1000-1300 $^{\circ}$C; the reference perovskites SrRhO$_3$ and SrIrO$_3$ were obtained at 8 GPa, 1200 $^{\circ}$C  and 6 GPa, 1000 $^{\circ}$C, respectively. About 30 wt.\% KClO$_4$ acting as oxidizing agent were added for synthesizing the compounds containing Co and Rh. The resultant KCl was washed away with deionized water.   Phase purity of the above samples was examined with powder X-ray diffraction (XRD) at room temperature with a Philips XÕpert diffractometer (Cu K$_{\alpha}$ radiation). All the A-site-ordered perovskites adopt a cubic \textit{IM$\overline{3}$} structure with lattice parameter increasing progressively from a = 7.1259(3) \AA\emph{} for M = Co, to a = 7.3933(1) \AA\emph{} for M = Rh, and to a = 7.4738(1) \AA\emph{} for M = Ir. On the other hand, the reference perovskites crystallize into the orthorhombic Pbnm structure with unit-cell parameters a = 5.5673(1) \AA\emph{}, b = 5.5399(2) \AA\emph{} and c = 7.8550(2) \AA\emph{} for SrRhO$_3$, and a = 5.5979(1) \AA\emph{}, b = 5.5669(1) \AA\emph{}, and c = 7.8909(1) \AA\emph{} for SrIrO$_3$, respectively.  

XAS measurements were carried out on the polycrystalline samples in the soft x-ray branch at the 4-ID-C beamline in the bulk-sensitive total fluorescence yield (TFY) and total electron yield (TEY) modes, with a 0.1 eV (0.3 eV) resolution at the O K-edge (Cu L-edge), at the Advanced Photon Source in Argonne National Laboratory. Measurements were taken on the Cu L-edge and O-K edge for all samples,  and all measurements shown here were obtained in TEY mode (TFY available in supplementary Fig. 1). To measure the 4$d$ and 5$d$ $B$-site valences, hard XAS measurements with a 1.5 (3) eV resolution were taken at the 4-ID-D beamline in transmission (fluorescence) mode for Ir (Rh). The ab-initio calculations were carried out in terms of density functional theory (DFT) within the generalized gradient approximation with supplemented Hubbard U (GGA+U) in plane wave as well as linear augmented plane wave basis. For the 4$d$ Rh and 5$d$ Ir compounds, calculations were carried out including spin-orbit (SO) coupling. U values were chosen to be 5 eV, 4 eV and 2 eV for Cu, Co and Rh/Ir respectively with Hund's exchange, J$_H$ chosen as 0.8 eV. The variation in chosen U values were checked.

In the first-principles DFT calculations we have primarily used the plane wave basis set and pseudo-potentials as implemented in the Vienna Ab-initio Simulation Package (VASP) \cite{vasp}.  The exchange-correlation function was chosen to be that of the generalized gradient approximation (GGA) implemented following the parametrization of Perdew-Burke-Ernzerhof \cite{PBE}.  The electron-electron correlation beyond GGA was taken into account through improved treatment of GGA + $U$ calculation within the + $U$ implementation of Dudarev {\it et al.}\cite{dudarev}. For the plane wave based calculations, we used projector augmented wave (PAW)\cite{paw} potentials. The wave functions were expanded in the plane wave basis with a kinetic energy cutoff of 600 eV and Brillouin zone summations were carried out with a 6 $\times$ 6 $\times$ 6 k-mesh. A $U$ value of 5 eV on Cu site was used. For the U value on the $B$ site a value of 4 eV was used for the 3d element Co and 1-2 eV was used for 4 d and 5 d elements, Rh and Ir.  The obtained results were verified in terms of variation of $U$ parameter. The Hund's rule coupling \textit{J} was fixed to 0.8 eV. The plane wave results were verified in terms of full-potential linearized augmented plane-wave (FLAPW) method as implemented\cite{wien2k} in WIEN2k. For FLPAW calculations, we chose the APW + lo as the basis set and the expansion in spherical harmonics for the radial wave functions was taken up to $l$ = 10. The charge densities and potentials were represented by spherical harmonics up to $l$ = 6. For Brillouin zone (BZ) integration, we considered about 200 k-points in the irreducible BZ and modified tetrahedron method was applied\cite{tetra} . The commonly used criterion for the convergence of basis sets relating the plane wave cutoff, K$_{max}$, and the smallest atomic sphere radius, R$_{MT}$, R$_{MT}$*K$_{max}$, was chosen to be 7.0. Spin-orbit coupling has been included in the calculations in scalar relativistic form as a perturbation to the original Hamiltonian.

In order to estimate the positions of the Cu d, $B$ d and O p energy levels as well as the plots of the effective Wannier functions for $B$ d states, we used  muffin-tin orbital (MTO) based N-th order MTO (NMTO)\cite{nmto}-downfolding calculations. Starting from full DFT calculations, NMTO-downfolding arrives at a few-orbital Hamiltonian by integrating out degrees which are not of interest. It does so by defining energy-selected, effective orbitals which serve as Wannier-like orbitals defining the few-orbital Hamiltonian in the {\it downfolded} representation. NMTO technique which is not yet available in its self-consistent form relies on the self-consistent potential parameters obtained out of linear muffin-tine orbital (LMTO)\cite{lmto} calculations. The results were cross-checked among the calculations in three different basis sets in terms of total energy differences, density of states and band structures.

\section{Acknowledgements}

JC is supported by  DOD-ARO Grant No. 0402-17291. JSZ and JBG is supported by NSF Grant. No. DMR-1122603. TS-D would also like to thank, CSIR and DST, India for funding. Work at the Advanced Photon Source, Argonne is supported by the U.S. Department of Energy, Office of Science under Grant No. DEAC02-06CH11357. JGC acknowledges the support from NSFC and MOST (Grant Nos. 11304371, 2014CB921500). We thank Dr. Shalinee Chikara and Prof. Gang Cao for sharing data on reference samples Rh$_2$O$_3$ and Sr$_2$RhO$_4$. JC acknowledges useful discussions with D. Khomskii.

\section{Contributions}
D. M., S.Middey, Y. Choi, D. H., B. A. G., J. W. F., and J. C. acquired the experimental data. S. Mukherjee and T. S-D. did the theoretical calculations. J. G. C., J. S. Z., and J. B. G. grew the samples. D.M and S. Middey analyzed data. All authors discussed the results. D.M., S. Middey, Y. Cao., T.S-D, J. C. wrote the manuscript.

\section{Competing financial interests}
The authors declare there are no competing financial interests.

 \clearpage

\begin{figure}[t!]
\vspace{-0pt}
\includegraphics[width=.45\textwidth]{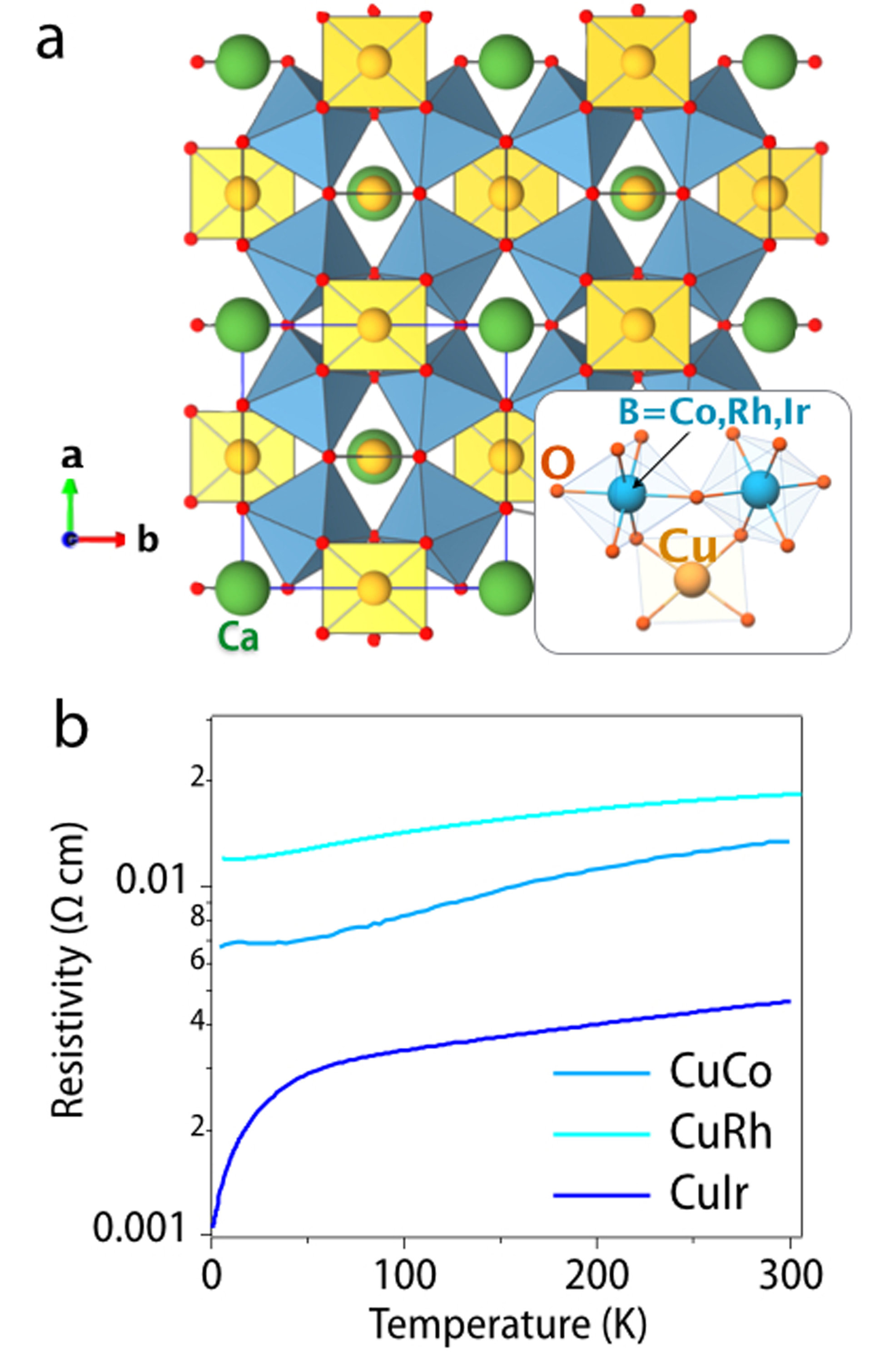}
\caption{\label{} \textbf{Structure and Properties of A-site ordered perovskites} a) Crystal structure of $A$-site ordered perovskites. Connection between CuO$_4$ planes and IrO$_6$ octahedra shown in the bottom right corner. b) Temperature dependent transport data for the $B$ = Co, Rh, Ir samples displaying the anomalous behavior for CCCIrO. CCCoO data taken from Ref. 22\cite{Yamada}. CCIrO data taken from Ref. 27\cite{ChengUn}.}
\end{figure}

\begin{figure}[t!]
\vspace{-0pt}
\includegraphics[width=.5\textwidth]{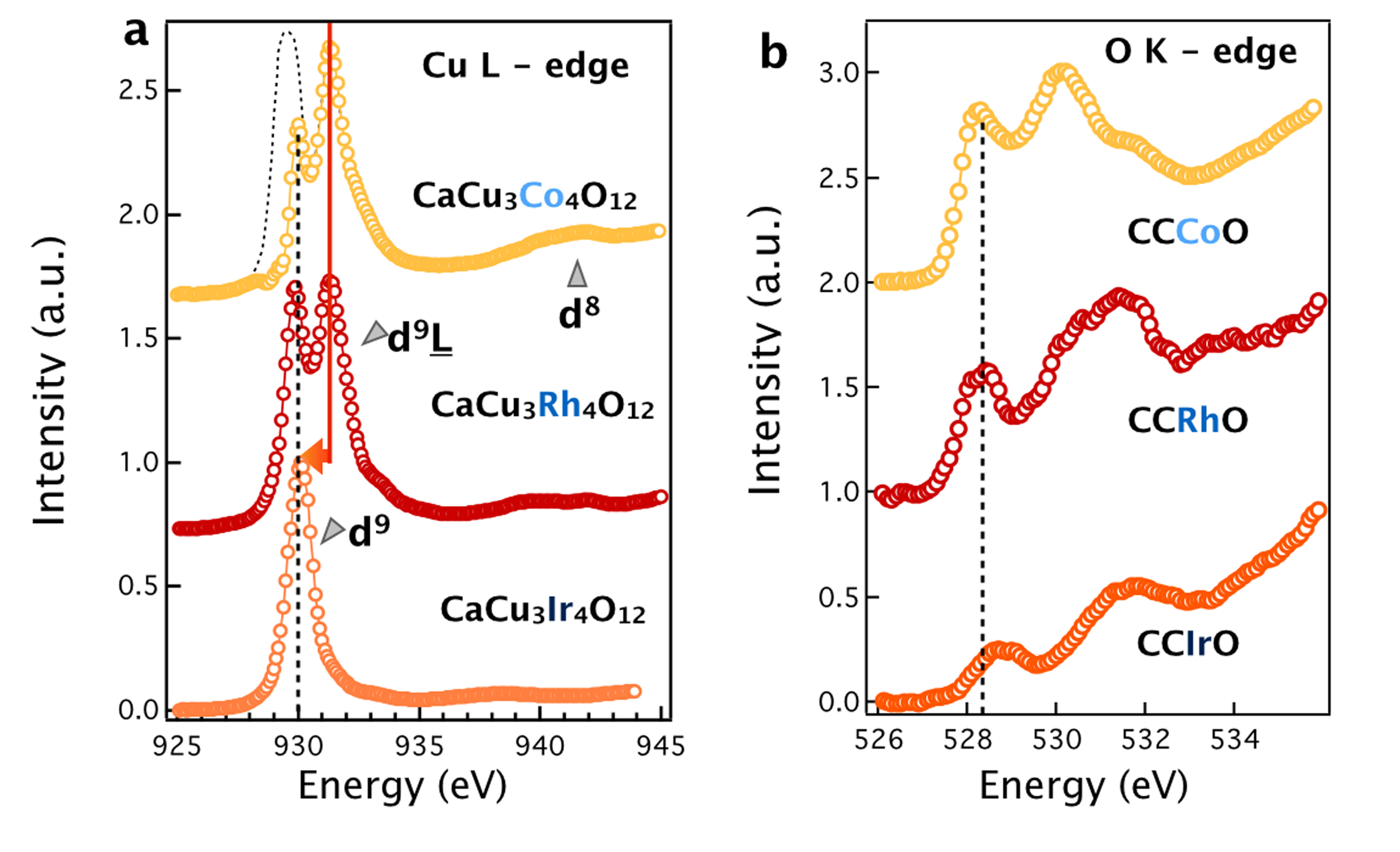}
\caption{\label{} \textbf{Changing Cu and O valency as a function of B-site.}
a) Soft XAS on the Cu L-edge for all samples showing the changing Cu valence. The dotted line is the spectrum before the subtraction of the impurity peak (Note that CCCoO TFY data and explanation of impurity peak originally given in Ref. 24\cite{Meyers}). The dashed line indicated the energy of the d$^9$ peak. b) XAS on the O K-edge showing both the reduction of the prepeak on O and the shifting of the O 2$p$ - Cu 3$d$ and O 2$p$ - $B$-site $d$ hybridized orbitals. The dashed line here indicates the energy of the O prepeak associated with a doped hole.}
\end{figure}

\begin{figure}[h]\vspace{-0pt}
\includegraphics[width=.5\textwidth]{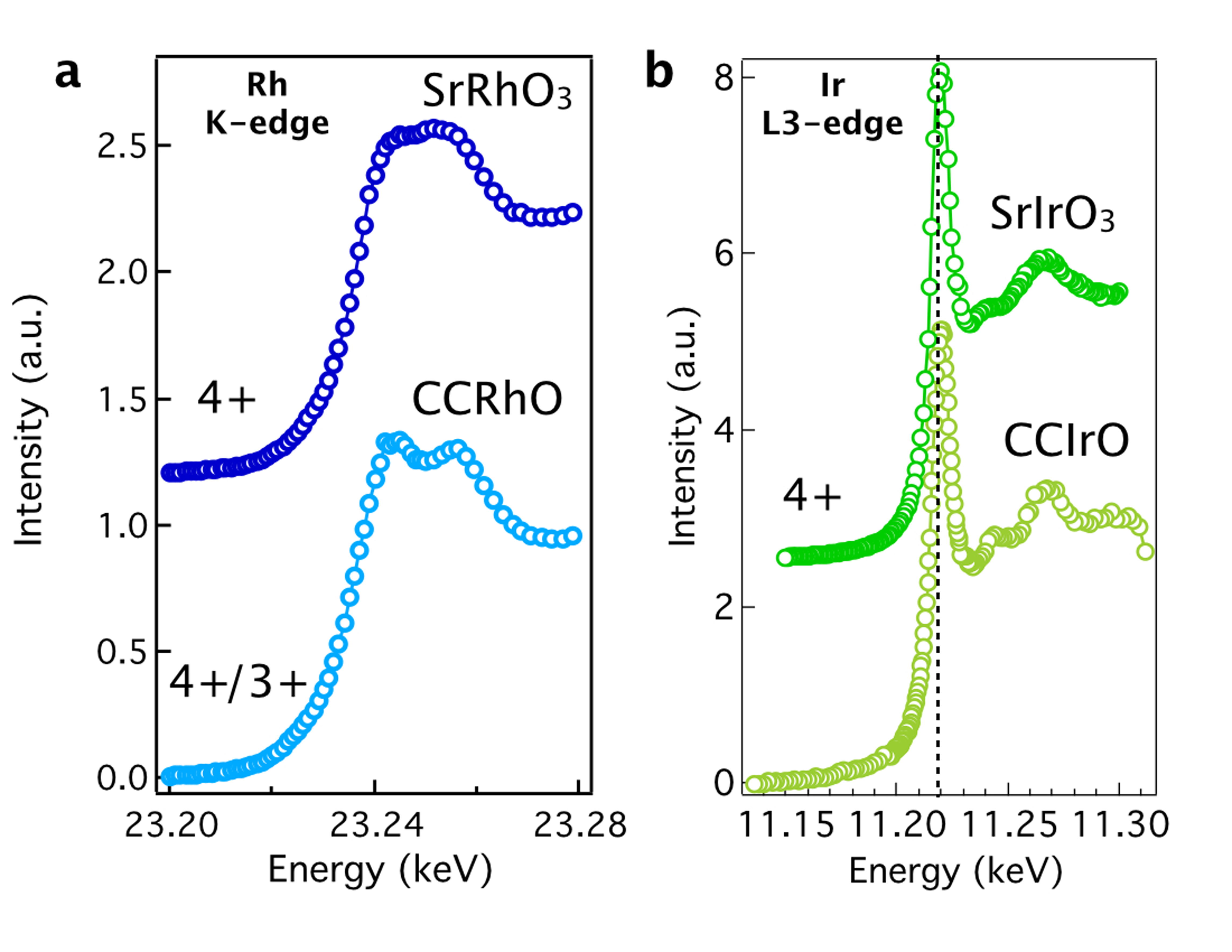}
\caption{\label{}   \textbf{Changing valency of the B-sites by hard XAS}
a) Hard XAS Rh K-edge measurements on the CCRhO
and SrRhO$_3$ (4+) standard. The line is a guide to the eyes.
b) XAS on the Ir L$_3$-edge for both CCIrO and a SrIrO$_3$
standard evidencing the nearly identical line shape and position indicating
the 4+ valency. The dashed line shows the excellent agreement of the peak positions.}
\end{figure}

\begin{figure}[t!]
\vspace{-0pt}
\includegraphics[width=.5\textwidth]{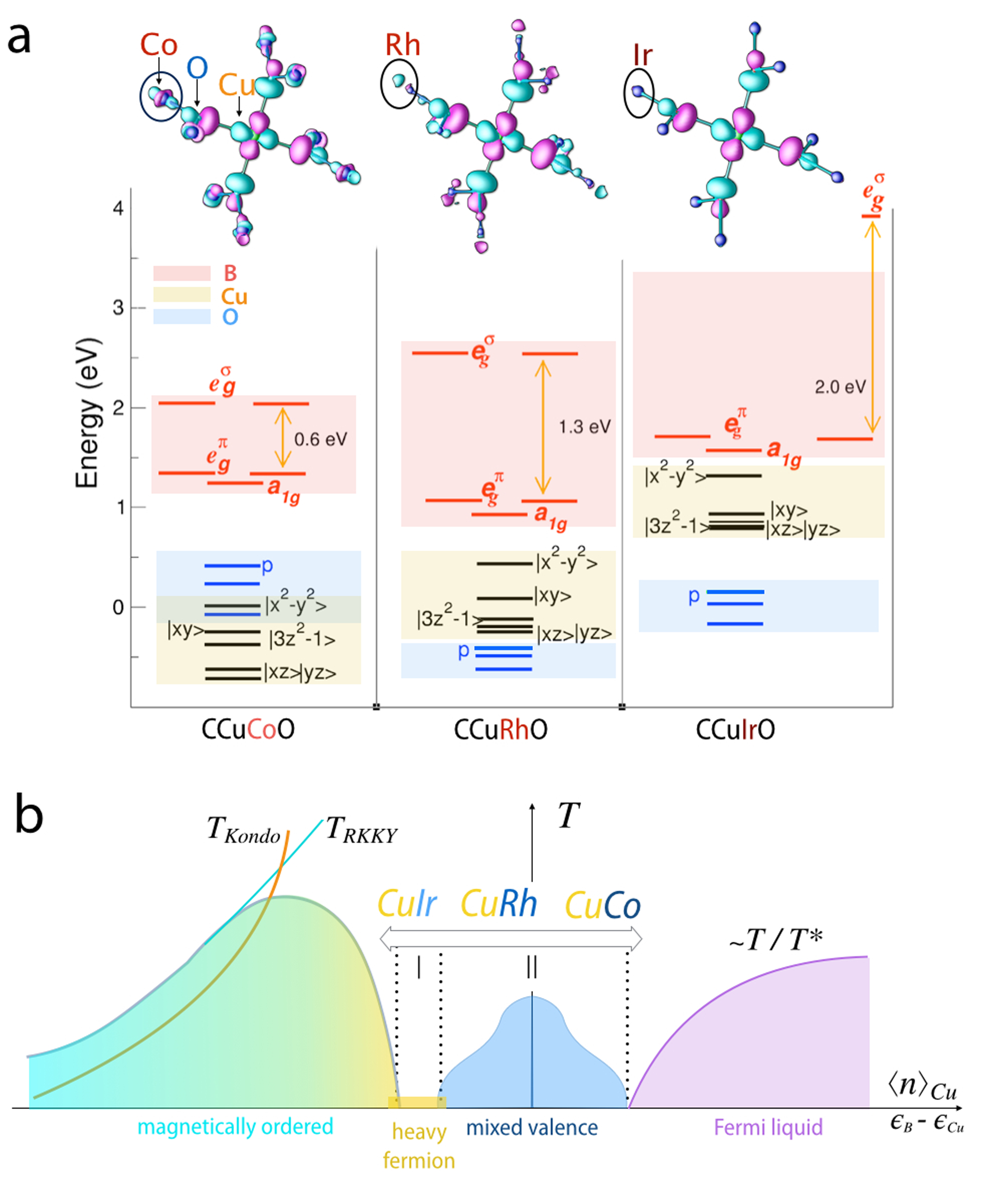}
\caption{\label{}   \textbf{Theoretical calculations and Doniach phase diagram.}
 a) Top panels: Plots of effective Wannier functions
for O p, Cu d$_{x^{2}-y^{2}}$ orbitals for CCCoO, CCRhO and CCIrO.
Plotted are the constant value surfaces with lobes of different signs colored
as cyan and magenta. The Cu, $B$- and O sites are shown
as green, red and blue colored balls. Bottom panels: Energy level positions
of Cu d, $B$ d and O p states for CCCoO, CCRhO and CCIrO. b) Doniach phase diagram showing the dependency
on the Cu occupation.}
\end{figure}

 \clearpage
\begin{table}[t!]
\caption{\label{} Experimental and theoretical values for the valence of Cu and the $B$-site for each compound}
\begin{center}
    \begin{tabular}{ | l | l | l | p{2.2cm} |}
    \hline
     & CCCoO & CCRhO & CCIrO \\ \hline
    Cu (experimental) & (2.90 $\pm$ 0.15)$^+$  & (2.63 $\pm$ 0.1)$^+$ & (2.03 $\pm$ 0.1)$^+$\\ \hline
    $B$-site (experimental) & 3.25$^+$ (ref. 24) & (3.64 $\pm$ 0.1)$^+$ & 4$^+$ \\ \hline
    Cu (theory) & 3$^+$ & 2.5$^+$ & 2$^+$\\ \hline
     $B$-site (theory) & 3.25$^+$ & 3.6$^+$ & 4$^+$ \\ \hline
     
    \end{tabular}
    
\end{center}
\end{table}

\end{document}